\documentclass[11pt,reqno]{amsart}
\usepackage{amsfonts, amssymb, amsmath, dsfont, graphicx, subcaption, enumerate, fullpage, xcolor}
\usepackage{algorithm}
\usepackage[noend]{algpseudocode}
\numberwithin{equation}{section}

\usepackage[capitalise]{cleveref}
\Crefname{subsection}{Subsection}{Subsections}
\crefname{subsection}{Subsection}{Subsections}
\crefname{line}{Line}{Lines}

\DeclareMathOperator{\argmin}{argmin}

\DeclareMathOperator{\Dirichlet}{Dirichlet}

\newcommand{\R}{\mathbb{R}}

\newcommand{\mX}{{\bf X}}
\newcommand{\mB}{{\bf B}}
\newcommand{\mA}{{\bf A}}
\newcommand{\mC}{{\bf C}}
\newcommand{\mY}{{\bf Y}}
\newcommand{\mW}{{\bf W}}
\newcommand{\mH}{{\bf H}}
\newcommand{\mU}{{\bf U}}
\newcommand{\cX}{{\mathcal X}}
\newcommand{\cA}{{\mathcal A}}
\newcommand{\cB}{{\mathcal B}}

\newcommand{\had}{\odot}

\newcommand{\thup}{^{\text{th}}}

\theoremstyle{remark}

\begin{document}

\bibliographystyle{abbrv}

\title{Sparseness-constrained Nonnegative Tensor Factorization
for Detecting Topics at Different Time Scales}
\author{Lara Kassab\,$^{1,*}$, Alona Kryshchenko\,$^{2}$, Hanbaek Lyu\,$^{3}$, Denali Molitor\,$^{1}$, Deanna Needell\,$^{1}$, Elizaveta Rebrova\,$^{4}$, Jiahong Yuan\,$^{3}$}

\address{$^{1}$Department of Mathematics, University of California, Los Angeles \\
$^{2}$Department of Mathematics, California State University Channel Islands \\
$^{3}$Department of Mathematics, University of Wisconsin - Madison \\ 
$^{4}$Department of Operations Research and Financial Engineering, Princeton University}
\email{$^{1}$lkassab@math.ucla.edu, deanna@math.ucla.edu, $^{2}$alona.kryshchenko@csuci.edu, $^{3}$hlyu@math.wisc.edu, jyuan93@wisc.edu, $^{4}$elre@princeton.edu}

\maketitle

\begin{abstract}
Temporal data (such as news articles or Twitter feeds) often consists of a mixture of long-lasting trends and popular but short-lasting topics of interest. A truly successful topic modeling strategy should be able to detect both types of topics and clearly locate them in time. 
    In this paper, we first show that nonnegative CANDECOMP/PARAFAC decomposition (NCPD) is able to discover topics of variable persistence automatically. Then, we propose sparseness-constrained NCPD (S-NCPD) and its online variant in order to actively control the length of the learned topics effectively and efficiently. Further, we propose quantitative ways to measure the topic length and demonstrate the ability of S-NCPD (as well as its online variant) to discover short and long-lasting temporal topics in a controlled manner in semi-synthetic and real-world data including news headlines. We also demonstrate that the online variant of S-NCPD reduces the reconstruction error more rapidly than S-NCPD. 
\end{abstract}

\section{Introduction}
\emph{Dynamic topic modeling} investigates how latent themes emerge, evolve, and fade in temporal text datasets.
Several works have examined topics and their evolution through time \cite{blei2006dynamic,hu2015modeling,iwata2010online,saha2012learning} using probabilistic models \cite{blei2006dynamic,wang2012continuous}, nonnegative matrix factorizations \cite{greene2017exploring,belford2016ensemble,cichocki2007nonnegative}, and deep learning models \cite{pathak2019adaptive}. For large and noisy datasets, such as social networks or news feed datasets, for the sake of interpretability, topic modeling techniques do not aim to recover all the topics, but only a subset of important topics. This raises a question of topic selection: What do we view as important? Motivated by this perspective, we propose dynamic topic modeling methods that can influence \emph{what kind of topics we recover}.
While some major topics may persist for an extended period of time, detecting short-lasting topics, that correspond to shorter-lasting, but impactful events or discussions, or seasonally trending periodic topics, could be as important. In this paper, we show that tensor-based methods are able to discover topics of variable persistence automatically. Moreover, we propose and compare two approaches to control the length of discovered topics, based on data chunking and on sparse decompositions.
We construct a semi-synthetic dataset based on the 20 Newsgroups dataset \cite{20news} to serve as a simple and well-understood experiment and real-world data based on the ABC news headlines dataset \cite{DVN/SYBGZL_2018}.

The two most popular classic techniques for topic modeling are Latent Dirichlet Allocation (LDA) \cite{blei2003latent} and Nonnegative Matrix Factorization (NMF) \cite{lee1999learning, lee2000algorithms}.
In LDA, one models a topic by a probability distribution on the set of words, which are evolved according to a Bayesian scheme by feeding in the batches of a $(\mathtt{words}\times \mathtt{documents})$ matrix to receive $(\mathtt{words}\times \mathtt{topics})$ and $(\mathtt{topics}\times \mathtt{documents})$ representations \cite{blei2006dynamic,hoffman2010online}. NMF is also a matrix-based method which decomposes the $(\mathtt{words}\times \mathtt{documents})$ matrix into $(\mathtt{words}\times \mathtt{topics})$ and $(\mathtt{topics}\times \mathtt{documents})$ matrices. When the documents have timestamps, that is, ordered in time, the $(\mathtt{topics}\times \mathtt{documents})$ matrix provides temporal ordering to the automatically detected topics.
However, one can note that given a large amount of time-stamped documents, such as news articles or tweets, topic evolution frequently happens not from one document to the next in time, but rather from a batch of nearly simultaneous documents to the next. 
For example, two consecutive tweets coming from two different users likely have no relation to each other. This suggests naturally multi-order, or tensorial, structure of large streams of temporal data. 

Tensor decompositions have many applications in machine learning~\cite{kolda2009tensor,rabanser2017introduction} including temporal analysis such as discovering
patterns~\cite{xiong2010temporal}, discovering time-evolving topics~\cite{bahargam2018constrained,bader2008discussion}, predicting evolution~\cite{dunlavy2011temporal} and more. Here, we focus on one of the most natural low-rank tensor decompositions based on CP-tensor rank, see, e.g.,  \cite{kolda2009tensor}. Recent prior work successfully employed nonnegative CP tensor decomposition for the discovery of temporal topics in text data \cite{ahn2020large,lyu2022online}. 
One can encode the entire corpus of documents as a \emph{3-dimensional tensor} where 
the three modes correspond to \emph{words}, \emph{relatively simultaneous documents}, and \emph{time}, respectively. This way, the time dimension of the tensor is designed to focus on temporal changes in the aggregated information from one-time slice to the next.

We believe the role of nonnegativity constraint on the temporal mode is crucial for the NCPD-type methods to be able to detect both long-lasting and short-lasting topics. 
Indeed, NMF is well-known to be able to extract spatially localized features when applied to image data \cite{lee1999learning} by using nonnegativity constraints on the spatial mode. Being a 3D analog of NMF, NCPD should be able to extract spatio-temporally localized features, which correspond to `short-lasting' (temporally localized) `topics' (spatially localized features) in our context of dynamic topic modeling. 
While nonnegative factorizations are used ubiquitously for topic identification and interpretability, there is less work that makes use of NCPD for this purpose, especially in terms of localization in the time domain, making it ever more important to study the differences in output when using a matrix versus a tensor factorization method with temporal data.

To our knowledge, existing methods typically do not study the prevalence of topics through time but rather how the topics themselves evolve through time (which is a very interesting problem on its own, e.g., \cite{blei2006dynamic,wang2012continuous}).
Furthermore, there is a lack of analysis of detecting short-lasting topics, proposing parameter choices for such detection (e.g., \cite{bader2008discussion}), or developing a method for detecting topics of targeted temporal structures. 

\subsection{Contributions}
While we find that NCPD is able to learn topics of various temporal structures, there is no means to `control' the type of temporal structures of the topics that we desire to learn. To overcome this difficulty, we propose a new method of NCPD that forces one of the factor matrices (specifically, the $(\mathtt{time}\times \mathtt{topic})$ factor)  to have a prescribed level of sparseness of its columns. We call this method the sparsity-constrained NCPD (S-NCPD). We propose a block-coordinate-descent-type algorithm that approximately finds such decomposition. Furthermore, inspired by the online NCPD algorithm in \cite{lyu2022online}, we also propose an online version of the S-NCPD algorithm that iteratively factorizes a sequence of smaller tensors while enforcing the same sparseness constraint as in S-NCPD. Our algorithm for OS-NCPD follows the framework of stochastic regularized majorization-minimizaiton \cite{lyu2023stochastic}. We experimentally validate that the proposed methods can successfully detect topics of desired temporal structure in real-world dynamic text data. Our contributions are summarized below. 

\begin{itemize}
    \item We demonstrate that NCPD is able to automatically detect and accurately represent topics of variable persistence from temporal text data.

    \item We propose Sparsity-constrained NCPD (S-NCPD) that actively controls the persistence of topics through constraining the sparseness of the columns of the $(\mathtt{time}\times \mathtt{topic})$ factor matrix. 

    \item We also propose an online version of S-NCPD (OS-NCPD), which has the same ability to control the persistence of learned topics as S-NCPD but is computationally more efficient than S-NCPD.

    \item  We introduce  $\alpha$-effective length and normalized AUC metric for quantitative measures for topic lengths. Using these measures, we validate that S-NCPD and OS-NCPD successfully detect topics of desired persistence in real-world data.
\end{itemize}

\subsection{Related Work}

Several works have examined topics and their evolution through time using probabilistic models \cite{blei2006dynamic,wang2012continuous}, nonnegative matrix factorizations \cite{greene2017exploring,belford2016ensemble,cichocki2007nonnegative}, and deep learning models \cite{pathak2019adaptive}.
In \cite{blei2006dynamic},  Blei and Lafferty propose 
a family of probabilistic time series models to analyze the time evolution of topics
in large document collections.
The model assumes that a discrete-time state space model
governs the evolution of the natural parameters of the
multinomial distributions that represent the topics.
In \cite{wang2012continuous}, the authors propose a continuous time dynamic topic model which uses Brownian
motion to model latent topics through a sequential collection of documents, where a “topic” is a pattern of word use that is
expected to evolve over the course of the collection.
Neither paper studies the prevalence of topics through time provides analysis on detecting short-lasting topics or proposes parameter choices for such detection. We also note that one of the advantages of NCPD and NMF over existing LDA methods is that there are far fewer parameter choices involved in the modeling process. 

Tensor decomposition techniques have numerous applications in machine learning~\cite{kolda2009tensor,rabanser2017introduction} including temporal analysis such as discovering
patterns~\cite{xiong2010temporal}, discovering time-evolving topics~\cite{bahargam2018constrained,bader2008discussion}, predicting evolution~\cite{dunlavy2011temporal}, modeling the behaviors of drug-target-disease interactions \cite{chen2019modeling}, and spotting anomalies~\cite{papalexakis2014spotting}. More recent related work in the line of research includes \cite{balasubramaniam2021identifying,balasubramaniam2020understanding,yu2023generalized}.
However, there is a lack of analysis of detecting short-lasting topics or proposing parameter choices for such detection (e.g., \cite{bader2008discussion}).

In \cite{ahn2020large}, the authors demonstrate NCPD as a dynamic modeling technique where critical temporal information is preserved, and events such as topic evolution, emergence, and fading are significantly easier to identify compared to NMF-based methods.  
In this work, we thoroughly study the persistence of temporal topics extracted from NCPD and its online variant \cite{lyu2022online}
propose quantitative ways to
measure the topic length, and control the length of discovered topics by (online) NCPD.

\subsection{Preliminaries and Notation}
We denote vectors with lowercase letters $x$ with $x(k)$ denoting its $k^{\textup{th}}$ entry, matrices with uppercase boldface letters, $\mX$, and third-order tensors with uppercase calligraphic letters $\cX$.  
\emph{Tensors} are common algebraic representations for multidimensional arrays.
The \emph{order} of a tensor is the number of dimensions, which is also referred to as \emph{ways} or \emph{modes}~\cite{kolda2009tensor}. 
For a matrix $\mX$, the vector $x_k$ denotes its $k\thup$ column. 
We let $\|\cdot\|_F$ and $\|\cdot\|_1$ denote the entrywise Frobenius norm, and the entrywise $L_1$ norm respectively. 
The set of nonnegative real numbers $[0,\infty)$ is denoted $\R_{\ge 0}$.
We let $\otimes$ denote the outer product of two vectors. For tensors $\cA$ and $\cB$ of the same size, denote by $\cA\had \cB$ the Hadamard (pointwise) product. When $\mB$ is a matrix, for each $1\le j \le n$, we denote their $j$-mode product by $\cA\times_{j} \mB$. See \cite{kolda2009tensor} for an excellent survey of related definitions and tensor algorithms.

\subsection{Organization}
In Section \ref{sec:tensor_methods}, we first introduce standard dynamic topic modeling methods: latent Dirichlet allocation (LDA), nonnegative matrix factorization (NMF), and nonnegative CP tensor decomposition (NCPD). Then we introduce sparsity-constrained NCPD (S-NCPD) and online S-NCPD as well as algorithms for solving the corresponding optimization problems. In Section \ref{sec:topic-lengths}, we introduce quantitative measures of the topic length. 
In Section \ref{sec:exper}, we analyze the performance of various dynamic topic modeling methods, including existing ones and the two newly proposed ones. Our focus is on the type of temporal structures of the topics learned by each method. We use semi-synthetic and real datasets in our experiments. Lastly, we include some discussions regarding those techniques and their results.

\section{Materials and Methods}

\subsection{Tensor factorization methods for topic modeling}
\label{sec:tensor_methods}
In this section, we discuss NMF, NCPD, and an online version of NCPD.

\iffalse
\subsubsection{Nonnegative Matrix Factorization}\label{subsec:nmf}
Nonnegative Matrix Factorization (NMF) is a popular tool for extracting hidden themes from text data \cite{buciu2008non,kuang2015nonnegative}. 
For a data matrix $\mX \in \R_{\ge 0}^{m\times n}$, one learns a low-rank dictionary $\mW \in \R_{\ge 0}^{m\times r}$ and code matrix $\mH \in \R_{\ge 0}^{r\times n}$ that minimize $\norm{\mX - \mW \mH}_F^2$,
where $r>0$ is typically chosen such that $r<\min\{m,n\}$.
Suppose $m$ denotes the number of features (in our case unigrams and bigrams) and $n$ the number of documents, then
the dictionary matrix $\mW$ represents \emph{topics} in terms of the original features. 
Each column of the code matrix $\mH$ represents a data point as a linear combination of the dictionary elements with nonnegative coefficients. 
We use NMF to learn a dictionary $\mW$ from all data and analyze topic dynamics through changes in topic prevalence over time in the code matrices from each time slice. 

\fi

\subsubsection{Latent Dirichlet Allocation}
Latent Dirichlet Allocation (LDA)  is another popular tool for extracting hidden topics from text data \cite{blei2003latent}. LDA is a hierarchical Bayesian model, in which words and documents are modeled as a finite mixture over an underlying set of topics. 
For each topic $k$, let $\beta_k$ be a multinomial distribution over the vocabulary which is assumed to have been drawn from a Dirichlet distribution $\Dirichlet(\eta)$. For each document $d$, let $\theta_d$ be a distribution over topics that are assumed to have Dirichlet prior $\Dirichlet(\alpha)$. These prior distributions are assumed to be symmetric. %$\alpha$ and $\eta$.
LDA then updates the prior distributions of $\beta$ and $\theta$ and approximates posterior distributions. 
Two approaches are commonly used to approximate posterior distributions  Markov Chain Monte Carlo (MCMC) methods and variational inference.

In our experiments, we consider an LDA model that uses online variational inference \cite{hoffman2010online}.  The posterior distribution of $\beta$ is used to find word representation of each topic and the posterior distribution of $\theta$ gives the topic distribution for each document. To learn topic dynamics over time, we take the mean over topic distributions $\theta_i$ for all the documents in each time slice and present them as columns of the heatmaps (e.g. Fig \ref{fig:LDA_headlines}).

\subsubsection{Nonnegative CP Tensor Decomposition}

Nonnegative CP Tensor Decomposition (NCPD) is a tool for decomposing higher-dimensional
data tensors into interpretable lower-dimensional representations. 
NCPD factorizes a tensor into a sum of nonnegative component rank-one tensors, defined as outer products of nonnegative vectors~\cite{carroll1970analysis,harshman1970foundations}. For instance,  given a third-order tensor $\cX \in \R^{n_1 \times n_2 \times n_3}_{\ge 0}$ and a fixed integer $r>0$, the approximate NCPD of $\cX$ seeks matrices $\mA \in \mathbb{R}^{n_1 \times r}_{\ge 0}, \mB \in \mathbb{R}^{n_2 \times r}_{\ge 0}, \mC \in \mathbb{R}^{n_3 \times r}_{\ge 0}$, such that 
$
  \cX \approx \sum_{k= 1}^{r} a_k \otimes b_k \otimes c_k,
$
where the nonnegative vectors $a_k$, $b_k$, and $c_k$ are the columns of $\mA, \mB$, and $\mC$, respectively.
The matrices $\mA$, $\mB$, and $\mC$ are referred to as the NCPD \emph{factor matrices}.
Such factor matrices are found by solving the following minimization problem 
\begin{equation}\label{eq:NCPD_reconstruction}
     \underset{\mA\in \R_{\ge 0}^{n_{1}\times r},\, \mB\in \R_{\ge 0}^{n_{2}\times r},\, \mC\in \R_{\ge 0}^{n_{3}\times r}}{\argmin} \left( \ell(\cX; \mA, \mB, \mC):=\left \|\cX - \sum \limits_{k= 1}^{r} a_k \otimes b_k \otimes c_k \right \|_F \right). 
\end{equation}
NCPD for decomposing any $d$th order data tensor can be formulated similarly. \textit{Nonnegative Matrix Factorization} (NMF) is a special instance of NCPD for decomposing second-order tensor data, which is a popular tool for extracting hidden themes from text data \cite{buciu2008non,kuang2015nonnegative}. 

Note that \eqref{eq:NCPD_reconstruction} is a non-convex optimization problem, but the objective function $\ell$ in \eqref{eq:NCPD_reconstruction} is block multi-convex (i.e., convex in each factor matrix while the other two factors are held fixed). Leveraging this structure, many researchers proposed algorithms for solving \eqref{eq:NCPD_reconstruction} have the nature of block coordinate descent (BCD) \cite{bertsekas1999nonlinear,wright2015coordinate}, including the multiplicative update algorithm~\cite{shashua2005non}, alternating least squares \cite{carroll1970analysis,harshman1970foundations}. Recently, Lyu and Li showed that regularized versions of these algorithms converge to the set of stationary points and can produce an $\epsilon$-stationary point of the objective in \eqref{eq:NCPD_reconstruction} within $\widetilde{O}(\epsilon^{-2})$ iterations \cite{lyu2023block}.

NCPD is considered a topic modeling technique for tensor data that successfully showcases topic variation across all modes of the tensor (including temporal mode(s))~\cite{ahn2020large}. 
Namely, suppose we have a third-order tensor data $\cX\in \R_{\ge 0}^{n_{1}\times n_{2}\times n_{3}}$ where $n_{1}=\mathtt{words}$ denotes the number of words in the vocabulary, $n_{2}=\mathtt{batch}$ denotes the number of documents in a  time slice, and $n_{3}=\mathtt{time}$ denotes the number of time slices. Applying NCPD to the third-order tensor data $\cX$, we obtain three factor matrices $\mA,\mB$, and $\mC$ of shapes $(\mathtt{words}\times r)$, $(\mathtt{batch}\times r)$, and $(\mathtt{time}\times r)$, respectively, where $r=\mathtt{topics}$ equals the number of topics we seek to find. We will be the most interested in the factor matrices $\mA$ and $\mC$; the columns of $\mA$ give $r$ topics in the data whereas the corresponding columns of $\mC$ give how their prevalence evolves through time. The second-factor matrix $\mB$ gives information on specific groups of documents that contributed to each discovered topic, which is of less importance for our purpose of dynamic topic modeling. 

\subsubsection{Sparseness-constrained NCPD (S-NCPD)}
\label{sec:SNCPD}

In order to control the temporal prevalence of learned topics, we propose to restrict the structure of the $(\mathtt{time}\times r)$ factor matrix $\mathbf{C}$ in NCPD as defined in equation \eqref{eq:NCPD_reconstruction} so that its columns have a `prescribed value of sparseness'. For this, we use the following measure of the sparseness of a vector introduced in 
Hoyer~\cite{hoyer2004non} in the context of NMF:
\begin{equation}\label{eq:Hoyer's Sparsity}
     s(\textbf{x}) := \frac{\sqrt{n} - (\sum |x_i|)/\sqrt{\sum x_i^2}}{\sqrt{n} -1}. 
\end{equation}
For a `temporal sparseness level' $\rho\in (0,1)$, we propose the following \textit{sparseness-constrained NCPD} (SNCPD): 
\begin{equation}\label{eq:NCPD_reconstruction_hoyer}
     \underset{\substack{\mA\in  \R_{\ge 0}^{n_{1}\times r},\, \mB\in \R_{\ge 0}^{n_{2}\times r},\, \mC\in \R_{\ge 0}^{n_{3}\times r} \\ s(C_{1})=\dots=s(C_{r})=\rho} }{\argmin}  \ell(\cX; \mA, \mB, \mC)
\end{equation}
where $f$ is as in \eqref{eq:NCPD_reconstruction} and $C_{j}$ denotes the $j$th column of the $(\mathtt{time}\times r)$ matrix $\mC$. Note that  $C_{j}$ describes the time evolution of the prevalence of the $j$th topic represented by the $j$th column $\mA[:,j]$ of the $(\mathtt{words}\times r)$ matrix $\mA$. Thus, the additional sparsity constraint on the columns of $\mC$ in \eqref{eq:NCPD_reconstruction_hoyer} actively controls the types of topics: for large $\rho$ we seek long-lasting topics, and for small $\rho$ we prefer short-lasting topics. We remark that \eqref{eq:NCPD_reconstruction_hoyer} is a tensorial extension of Hoyer's sparsity-constrained NMF \cite{hoyer2004non}, where the goal is to control the sparsity of the dictionary atoms learned by NMF. Our insight is to use a similar technique to control the temporal signature of topics learned by NCPD.

Since \eqref{eq:NCPD_reconstruction_hoyer} is still a block multi-convex minimization problem, in order to compute  an approximate optimum  for the SNCPD problem, we may use a modified version of BCD of the following form:
\begin{align}\label{alg:SNCPD}
\begin{cases}
    \mA_{t} \leftarrow \underset{\mA\in  \R_{\ge 0}^{n_{1}\times r}}\argmin \ell(\cX; \mA, \mB_{t-1}, \mC_{t-1}), \qquad \mB_{t} \leftarrow \underset{\mB \in  \R_{\ge 0}^{n_{2}\times r}}\argmin \ell(\cX; \mA_{t}, \mB, \mC_{t-1}), \\
    \mC_{t} \leftarrow \underset{\substack{\mC \in  \R_{\ge 0}^{n_{3}\times r}  \\ s(C_{1})=\dots=s(C_{r})=\rho}   }{\argmin} \ell(\cX; \mA_{t}, \mB_{t}, \mC). 
\end{cases}
\end{align}
The constraint on the temporal factor $\mC_{t}$ in \eqref{alg:SNCPD} is given by the intersection of the nonnegativity and sparseness constraints. The latter is the set of all vectors in $\R^{n_{3}}$ with a fixed ratio between the $L_{1}$- and $L_{2}$-norms (depending on $\rho$), which is unfortunately not a convex constraint. Hence known theoretical results for BCD with convex constraints sets (e.g., \cite{lyu2023block}) do not apply, and we will need to compute an approximate solution $\hat{\mC}_{t}$ for $\mC_{t}$. In order to do this, we use the following projected-gradient-decent-type algorithm for sparseness-constrained nonnegative least squares:
\begin{algorithm}[H]
		\caption{Sparseness-constrained Nonnegative Least Squares (S-NLS)}
        \label{algorithm:SNLS}		
        \begin{algorithmic}
			\State \textbf{Input:} Matrix $\mY\in \R^{p\times n}_{\ge 0}$; Matrix $\mW\in \R^{p\times r}_{\ge 0}$;  Sparseness level $\rho\in (0,1)$; Iteration number $T$
			\State \textbf{output:} Approximate solution $\hat{\mH}$ for $\mH = \argmin_{\mU\in \R^{r\times n}_{\ge 0},\, s(U_{1})=\cdots=s(U_{1})=\rho} \lVert \mY - \mW \mH \rVert_{F}^{2}$
           
            \State \quad \textbf{For $t=1,\dots,T$}: 
            \State \qquad \textbf{For $i=1,\dots,r$}:  \hspace{7.92cm} ($\triangleright$ update rows of $\mH$ cyclically)
            %\State \quad \quad $A\leftarrow W^{T}W\in \R^{r\times r}$
            \State \quad \qquad $x \leftarrow \mH[i,:] - \frac{t^{-1}}{\mW^{T} \mW[i,i]+1} \mW^{T}(\mW \mH - \mY)$ \hspace{0.8cm} ($\triangleright$ gradient descent with an adaptive stepsize)
            \State \quad\qquad $x \leftarrow \textup{Sparsify}_{\rho}(x)$ \hspace{1cm} ($\triangleright$ Hoyer's alternating projection for sparsification, see \cite{hoyer2004non})
            \State \quad\qquad $\mH[i,:] \leftarrow \max(\mathbf{0}, x)$ \hspace{7.6cm} ($\triangleright$ nonnegativity projection)
            \State \quad\quad \textbf{End For} 
             \State \quad \textbf{End For} 
		\end{algorithmic}
	\end{algorithm}
In order to compute $\hat{\mC}_{t}$ in \eqref{alg:SNCPD}, we use  \cref{algorithm:SNLS} with $\mY\in \R^{n_{1}n_{2}\times n_{3}}$ the mode-3 unfolding of $\cX$ and $\mW\in \R^{n_{1}n_{2}\times r}$  whose columns are vectorization of the outer products of respective columns of $\mA_{t}$ and $\mB_{t}$. Hoyer's alternating projection for sparsification \cite{hoyer2004non} finds a  nearby vector that approximately matches the desired sparseness level. Note that high (resp., low) values of $\rho$ result in topics that have sparse (resp., dense) prevalence (e.g., columns of the $(\mathtt{time}\times r)$ factor matrix).

\subsubsection{Sparseness-constrained Online Nonnegative CP Decomposition (OS-NCPD)}\label{sec:ONCPD} 

The computational cost of applying S-NCPD to a large 3D tensor may be computationally infeasible. Following the Online NCPD by Lyu, Strohmeier, and Needell \cite{lyu2022online}, here we propose an online version of SNCPD that we call Online S-NCPD (OS-NCPD for short). This method is a mini-batch extension of the batch S-NCPD \eqref{eq:NCPD_reconstruction_hoyer}, where mini-batches of sub-3D tensors are processed in a sequential manner to progressively compute a $(\mathtt{words}\times r)$ factor $\mA$ and $(\mathtt{time}\times r)$ factor $\mC$ with column-wise sparseness constraint.  

The key idea behind OS-NCPD is as follows. 
Recall that each temporal slice of the 3D tensor consists of multiple `simultaneous' documents in the time domain. In our application, extracting features from a batch of documents coming from the same time slice is not of major importance. So, what if on each time slice we subsample only a small number $\mathtt{batch}'\ll \mathtt{batch}$ of documents, and apply S-NCPD to the resulting smaller tensor $\cX'$ of shape $(\mathtt{words}\times \mathtt{batch}'\times \mathtt{time})$? This will give us three factor matrices $\mA,\mB'$, and $\mC$ of shapes $(\mathtt{words}\times r)$, $(\mathtt{batch}'\times r)$, and $(\mathtt{time}\times r)$, respectively, where the first and last factor matrices $\mA$ and $\mC$ have the same shapes as before. While using S-NCPD on a single subsample of the original tensor $\cX$ has reduced computational cost, we may also lose some information since we only learn from a single subsample. However, it is possible to process a number of such subsamples in a sequential manner, so that each factorization problem has a reduced dimension but the factor matrices $\mA$ and $\mC$ improve over subsamples.

The OS-NCPD can be formulated by a stochastic program as follows. Given a probability distribution $\pi$ on the set of data tensors $\R_{\ge 0}^{n_{1}\times n_{2}'\times n_{3}}$, consider seeking 
  nonnegative factor matrices $\mathbf{A}\in \R_{\ge 0}^{n_{1}\times r}$ and $\mathbf{C}\in \R_{\ge 0}^{n_{3}\times r}$ by solving the following stochastic program
\begin{equation}\label{eq:expected_RE_S}
    \underset{\substack{\mA,\,  \mC \\ s(C_{1})=\dots=S(C_{r})=\rho }  }{\argmin} \,\,\mathbb{E}_{\cX\sim \pi}\left[ \underset{\mB\in \R_{\ge 0}^{n_{2}' \times r}}{\inf}  \ell(\cX; \mA,\mB, \mC)
    \right],
\end{equation}
where the \textit{random} data tensor $\cX$ is sampled from the distribution $\pi$.  The stochastic program \eqref{eq:expected_RE_S} is equivalent to the S-NCPD problem \eqref{eq:NCPD_reconstruction_hoyer} when the distribution $\pi$ is supported on a single data tensor. %Indeed, in this case, one can absorb $h(k)$ into $a_{k}$ as there is a single data tensor to factorize. 

We propose the following iterative algorithm for solving \eqref{eq:expected_RE_S}, which is a minor modification for the Online CP-dictionary learning (OCPDL) algorithm in \cite{lyu2022online}. Suppose we have learned the  loading matrices $\mathbf{A}_{t-1},\mathbf{C}_{t-1}$ from the sequence $\cX_{1},\dots,\cX_{t-1}$ of data tensors in $\R_{\ge 0}^{n_{1}\times n_{2}' \times n_{3}}$. Then we compute the updated loading matrices $[ \mathbf{A}_{t},\mathbf{C}_{t} ]$ by
	\begin{align}\label{alg:OS_NCPD}
		\begin{cases}
			\mB_t &\leftarrow {\color{black}\underset{\mB\in \R_{\ge 0}^{n_{2}'\times r}}{\argmin} \ell(\cX_{t}; \mB, \mathbf{A}_{t-1},\mathbf{C}_{t-1}) } \\
			\hat{f}_{t}(\mathbf{A},\mathbf{C}) &\leftarrow  (1-w_{t}) \hat{f}_{t-1}(\mathbf{A},\mathbf{C}) + w_{t}  \ell(\cX_{t}; \mathbf{A}, \mB_{t}, \mathbf{C}) \\[5pt]
			\mathbf{A}_{t} &\leftarrow  \underset{\mathbf{A}\in \R_{\ge 0}^{n_{1}\times r},\, \lVert \mathbf{A} - \mathbf{A}_{t-1}\rVert_{F}\le w_{t}}{\argmin} \,\,  \hat{f}_{t}(\mathbf{A},\mathbf{C}_{t-1})\\
			\mathbf{C}_{t} &\leftarrow  \underset{\substack{
            \mathbf{C}\in \R_{\ge 0}^{n_{3}\times r},\, \lVert \mathbf{C} - \mathbf{C}_{t-1}\rVert_{F}\le w_{t}
            \\ {\color{black}s(C_{1})=\dots=S(C_{r})=\rho}}}{\argmin} \,\,  \hat{f}_{t}(\mathbf{A}_{t}, \mathbf{C}),
		\end{cases}
	\end{align}
	where $\lambda\ge 0$ is an absolute constant and $(w_{t})_{t\ge 1}$ is a non-increasing sequence of weights in $(0,1]$. The recursively defined function $\hat{f}_{t}:(\mA,\mC) \mapsto [0,\infty)$ is called the \textit{surrogate loss function}, which is quadratic in each factor $\mA$ and $\mC$ but is not jointly convex. Namely, when the new tensor data $\cX_{t}$ arrives, one computes the $(\mathtt{batch}'\times r)$ factor $\mB_{t}\in \R^{n_{2}'\times r}_{\ge 0}$ for $\cX_{t}$ with respect to the previous loading matrices in $(\mA_{t-1}, \mC_{t-1})$, updates the surrogate loss function $\hat{f}_{t}$, and then \textit{sequentially} minimizes it to find updated loading matrices within diminishing search radius $w_{t}$. In our implementation, for each $t$, we subsample a tensor $\mathcal{X}_{t}$ of shape $n_{1}\times n_{2}\times n_{3}$ from $\mathcal{X}$ uniformly at random. During the execution of the algorithm, one only needs to store a matrix of dimension $n_{1} n_{2}' n_{3} r$, regardless of the total number of iteration $T$. We refer the reader to \cite{lyu2022online} for more details. 
 
    In comparison to the original OCPDL algorithm, in \eqref{alg:OS_NCPD} we added additional sparsity constraint on the columns of $\mC_{t}$. An approximate solution $\hat{\mC}_{t}$ for $\mC_{t}$ can be computed using a projected gradient descent method similar to Algorithm \ref{algorithm:SNLS}. The original OCPDL algorithm is guaranteed to almost surely converge to the set of stationary points of the objective of \eqref{eq:expected_RE_S} and shows a superior convergence rate against standard (offline) algorithms for NCPD. Recently in \cite{lyu2023stochastic}, it was shown that this algorithm can produce an $\epsilon$-approximate stationary point of the objective within $\widetilde{O}(\epsilon^{-4})$ iterations. 

\iffalse
\subsection{Temporal Chunking}
\label{ss:temporal_chunking}

We propose another method to control the temporal structure of topics learned by NCPD by a suitable preprocessing of the tensor data. For instance, suppose that the $i$th temporal slice of our tensor data $\mathcal{X}\in \R^{n_{1}\times n_{2}\times n_{3}}$ represents the documents we observe during the $i$th month. Fix an integer $b\ge 1$ and consider concatenating $b$ consecutive temporal slices of $\mathcal{X}$ to form a new tensor $\mathcal{X}_{b}$ so that its $j$th temporal slice is the concatenation of the $j$th $b$ consecutive temporal slices of $\mathcal{X}$ (of shape $n_{1}\times (n_{2} b)$). This corresponds to changing the unit of time from a month to $b$ consecutive months, and regarding all documents within the $j$th $b$ consecutive temporal slices in $\mathcal{X}$ as `simultaneous'. We call this preprocessing of the data tensor as \textit{temporal chunking}. 
\fi

\subsection{Quantifying Lengths of Topics}
\label{sec:topic-lengths}
How can we determine the ``length'' of a topic found by any of the described methods? 
How can we judge whether a topic is considered ``short-lasting" or "long-lasting"?

First, we can judge the topic lengths visually based on the heatmaps of matrix $\mathbf{T}\in \R_+^{r \times n}$ representing the dynamics of the topics over time where $r$ denotes the number of topics and $n$ number of time units or stamps. 
In the case of NCPD, S-NCPD, and OS-NCPD, $\mathbf{T} = \mathbf{C}$ is a temporal factor matrix, and in the cases of NMF and LDA, the columns of $\mathbf{T}$ are topic intensities over the time slices. 
By construction, this matrix $\mathbf{T}$ has normalized columns. 
Qualitatively, approximately sparse rows of the matrix $\mathbf{T}$ correspond to the topics that were trending shortly or periodically. 

To complement this qualitative analysis of the topics' lengths, in this section, we propose a metric to quantify the notion of the length of a topic. 
This way, one can explicitly parametrize the effective (approximate) length of each topic and demonstrate the variability of the topic lengths discovered by the tensor-based methods. 

Our proposed metric quantifies the number of consecutive time units required to cover a certain ``proportion" of the topic that we denote by $\alpha$.
We consider the matrix $\tilde{\mathbf{T}} \in \R_+^{r \times n}$ which is the matrix $\mathbf{T}$ with the rows normalized to add up to 1. 
Normalization of the rows produces a probability distribution for each individual topic over time. 
Informally, it captures how many consecutive time units are required for each topic to include a certain proportion of its whole ``mass". 
Specifically, for a fraction $\alpha \in [0,1]$ and the topic $\tau$, its \textit{$\alpha$-effective length} denoted by $\ell_{\alpha}(\tau)$, is defined as

\begin{equation}\label{eq:topic_length}
\ell_{\alpha}(\tau) := \min \limits_{i \in [n-1]} \bigg\{ l \quad \bigg| \quad \sum\limits_{j = i}^{i+l}\tilde{\mathbf{T}}[\tau, j] > \alpha \bigg\} 
\end{equation}

By definition, for $\alpha = 0$, all the topics will have zero length. For $\alpha = 1$, the length of the topic is the total number of nonzero entries in the corresponding row. Typically, the intermediate values of $\alpha$ could demonstrate the variability of the topic lengths. 
The choice of parameter $\alpha$ can be determined by a specific application. Visually, this technique acts as an ``elbow method'' as $\alpha$ varies, where we can also observe the re-occurrence of a topic by the number of elbows in the curve (e.g., Figure~\ref{fig:20news_lens}, NCPD, topic (3)).
Accordingly, we choose $\alpha = 60\%$ in Figure~\ref{fig:nAUC_plot}\textbf{a} for the news headlines dataset, and observe smaller mean and greater standard deviation for the $0.6$-effective lengths of the topics generated by NCPD and ONCPD.

By varying the value of $\alpha$ in $[0,1]$, one obtains plots of the function $\alpha\mapsto l_{\alpha}(\tau)$, which we refer to the \textit{topic ROC}, from which various information on temporal features of learned topics can be extracted. We note the following elementary but useful observations on topic ROCs:
\begin{enumerate}
    \item[(a)] The diagonal line in topic ROC corresponds to topics that are uniformly distributed over the entire time horizon;   
    \item[(b)] For any topic $\tau$, its topic ROC lies beneath the diagonal line;
    \item[(c)] If a topic $\tau$ is fully covered by $k$-consecutive time units, i.e., $\ell_{1}(\tau)=k$, then its topic ROC lies beneath the line segments from $(0,0)$ to $(1,k)$. 
\end{enumerate}

Based on the above observations, it is also possible to give a single \emph{persistence score} for each topic, that is, a number independent of other parameters (such as $\alpha$) and of visual judgment. One of many ways to define it is to aggregate the $\alpha$-effective lengths with various values of $\alpha$, measuring the area under the curve $\alpha\mapsto l_{\alpha}(\cdot)$, and normalizing it by $1/2$ total number of time slices in the time range. 
Such normalization guarantees all the persistence (nAUC) scores to be in the range between $0$ and $1$, since the curve $l_{\alpha}$ always lies under the diagonal.
Indeed, nAUC equals 1 corresponds to the ``most persistent" topic having equal weights at each time slice in the range (observation (a)). 
Further, if a topic is fully covered by a short time interval, its nAUC score would be close to 0 (observation (c)).

We note that multiple variations of the definition~\eqref{eq:topic_length} are possible and might be preferable in some applications. 
For example, the alternative measure that considers non-consecutive time unit contributions to the topic length would be able to detect periodic topics like the ones we can visually observe in topic 16 in Figure~\ref{fig:sncpd}.

\section{Results} \label{sec:exper}
In this section, we compare the performance of NMF, LDA, NCPD, and ONCPD methods in identifying temporal topics in semi-synthetic and real datasets. 

\subsection{Experimental Setup}
In all the experiments, documents are converted to 
term frequency-inverse document frequency (TFIDF) vector representations using the sklearn TFIDFVectorizer \cite{scikit-learn}.
We compute NMF of the data matrix using sklearn~\cite{scikit-learn} with nonnegative double singular value decomposition initialization~\cite{boutsidis2008svd}.
We compute NCPD of the tensor data with multiplicative updates \cite{shashua2005non} using TensorLy~\cite{tensorly} and SVD initialization.
We compute ONCPD using the Online CP-Dictionary Learning algorithm in  \cite{lyu2022online} with SVD initialization. The subsampled batch size ($\mathtt{batch}' = n_{2}'$) for ONCPD (see Section \ref{sec:ONCPD}) equals $5$ for 20 Newsgroups (full $\mathtt{batch} = n_{2}=26$, see Section \ref{subsec:semi synth}) and $100$ for the Headlines dataset ($n_{2}=700$, see Section \ref{subsec:headlines exp}). For S-NCPD and OS-NCPD, we implemented algorithms \eqref{alg:SNCPD} and \eqref{alg:OS_NCPD}, respectively. Lastly, for LDA we construct a bag-of-words corpus using the same dictionary as the other methods (obtained from the TFIDF weights) and compute the model using gensim LDA model~\cite{rehurek_lrec} with various numbers of passes and training chunks to save memory on larger datasets~\cite{hoffman2010online}. 

The keyword representation of each of the extracted topics is also provided for interpretability. Each learned topic is represented by a positive linear combination of terms. Terms with larger values in a particular topic are more significant for that topic and, thus, the terms with the largest values provide interpretable descriptions of the topics.
The number of topics for the synthetic 20 Newsgroups dataset is chosen to match the known number of article subjects. 
For complex real-world data, News Headlines datasets, we choose the number of topics to balance readability and the discovery of relevant events. We believe that increasing the number of topics could reveal additional relevant topics. 

\subsection{Semi-synthetic Dynamic Dataset Results}\label{subsec:semi synth}
The 20 Newsgroups dataset~\cite{20news} is a collection of documents divided into six groups partitioned into subjects, with a total of 20 subtopics. This dataset is commonly used as an experimental benchmark for document classification and
clustering.
We consider a semi-synthetic dataset constructed from the 20 Newsgroups dataset to illustrate the dynamic topic modeling performance of NMF, LDA, NCPD, and ONCPD on a simple and well-understood dataset. 
\begin{figure}[ht!]
    \centering
    \includegraphics[scale=0.25]{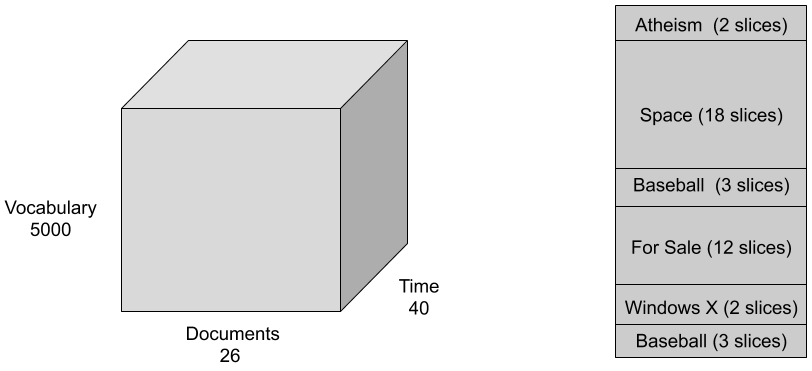}
    \caption{Semi-synthetic 20 Newsgroups tensor construction.}
    \label{fig:20news_dataset_construction}
\end{figure}

We consider only five categories: ``Atheism", ``Space", ``Baseball", ``For Sale", and ``Windows X" with a total of 1040 documents. 
We remove headers, footers, and quotes from all documents and compute TFIDF representation with a vocabulary size equal to 5000.
The NLTK English stopword list~\cite{nltk}, and words appearing in more than 95\% of the documents are removed.
We organize the dataset into a $5000 \times 26 \times 40$ tensor with dimensions: vocabulary size by number of documents by time.
Each time slice consists entirely of articles from the same category, and the categories of the times slices are ordered as: (``Aethism", time slices 1-2), (``Space", time slices 3-20), (``Baseball", time slices 21-23), (``For Sale", time slices 24-35), (``Windows X", time slices 36-37), and (``Baseball", time slices 37-40).
The tensor is illustrated in \Cref{fig:20news_dataset_construction}.
We run NMF, LDA, NCPD, and ONCPD as described in \Cref{sec:tensor_methods} with a rank equal to 5 reflecting the number of categories in the dataset. 
In this section, for NMF and LDA, we first unfold the tensor along the time mode, learn the topics, and then compute the mean topic representation for each time slice.
\begin{figure*}[!t]
    \centering
    \small
     \begin{tabular}{p{0.02\textwidth}|p{0.18\textwidth}|p{0.21\textwidth}|p{0.26\textwidth}|p{0.28\textwidth}}
     \noindent

         \textbf{} & \quad\quad\quad\textbf{LDA} & \quad\quad\quad\textbf{NMF} & \quad\quad\quad\textbf{NCPD} &
         \quad\quad\quad\textbf{ONCPD} 
         \\ \hline
         (1) &  would, like, one \quad &  \quad would, like, think\quad  & space, would, like \quad & 
         space, would, nasa 
         \\ 
         (2) &  edu, use, window  \quad & \quad drive, sale, offer
          \quad & 00, sale, drive  \quad &   00, sale, drive 
         \\ 
         (3) &  space, launch, nasa \quad &  \quad space, shuttle, nasa \quad  &  games, game, year \quad &  games, year, game
         \\ 
         (4) & new, sale, please \quad\quad & \quad  00, 20, 50\quad  &  god, believe, religion \quad &  god, wrong, people 
         \\ 
         (5) & 00, 50, 20  \quad\quad &  \quad mac, hm, msu\quad & window, widget, application  \quad &   window, widget, application          
     \end{tabular}
     \\
     \quad \includegraphics[scale =0.28]{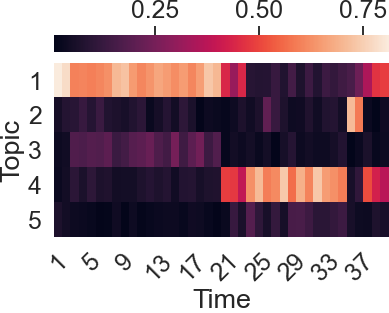}  
     \quad \includegraphics[scale =0.28]{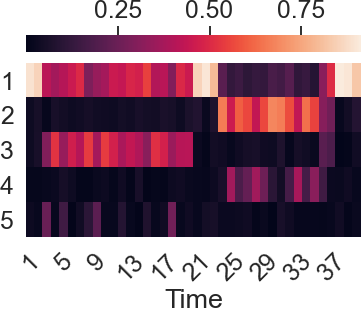}
     \quad \includegraphics[scale =0.28]{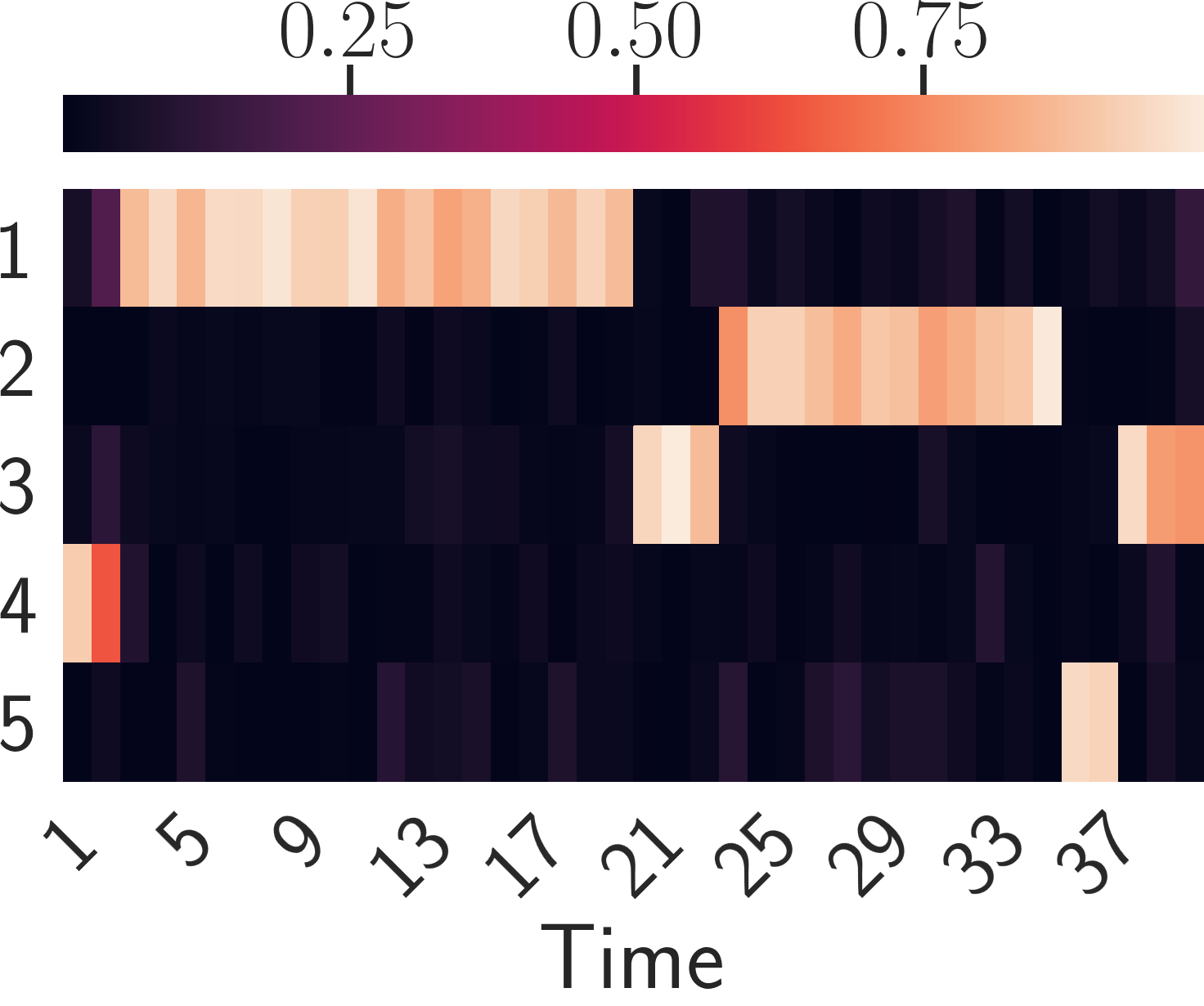}
     \quad \includegraphics[scale =0.28]{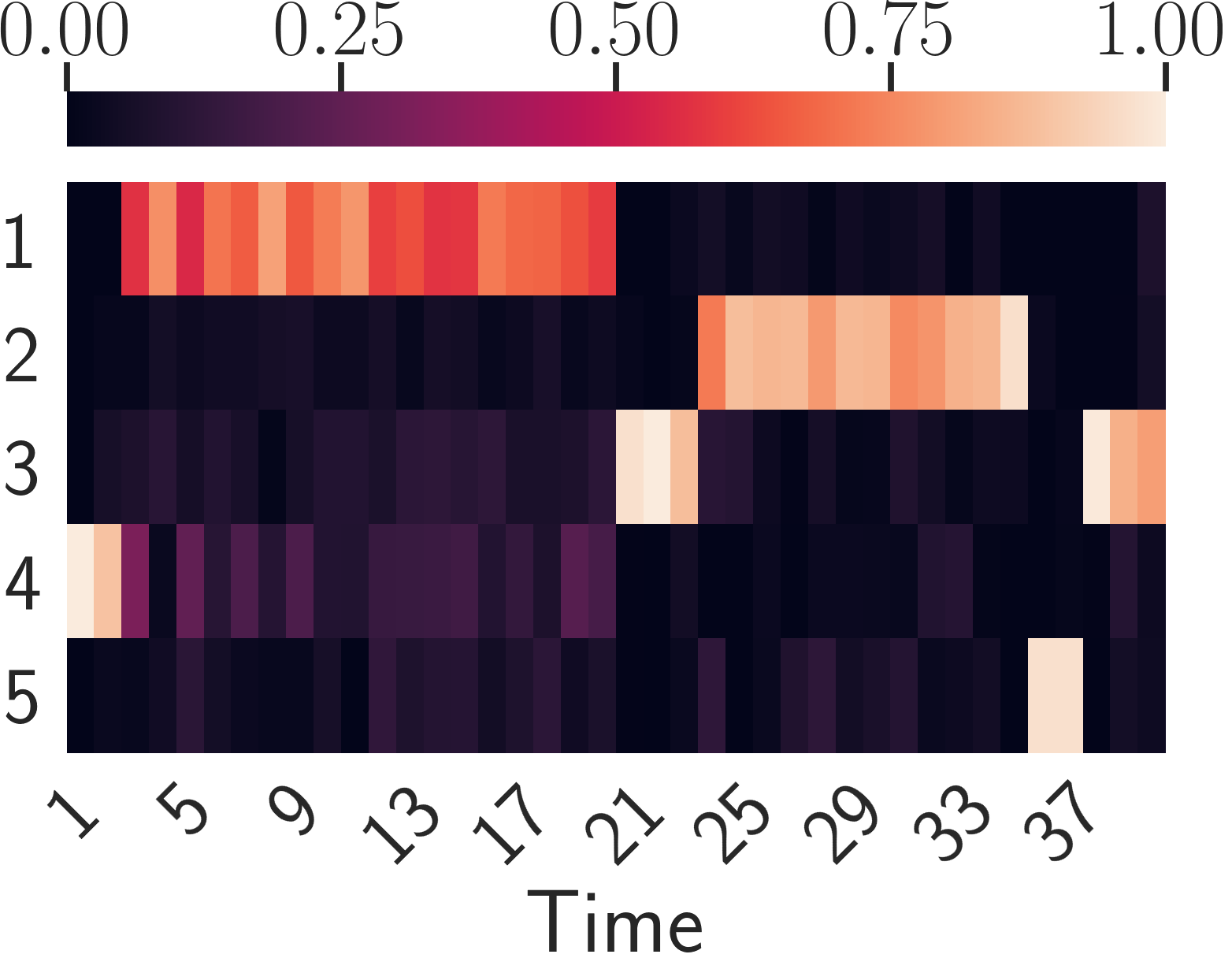} 
     \\
    \caption{The learned topics and prevalence of each extracted topic from the semi-synthetic 20 Newsgroups dataset are shown for each of the four models (LDA, NMF, NCPD, ONCPD). 
    The columns of each heatmap indicate the distribution of the extracted topics for each time slice. The top three keywords corresponding to each topic of the models are provided in the table.}
    \label{fig:20news_semi-synthetic_heatmaps}
\end{figure*}

Learned topics and the prevalence of each topic over time are indicated for each method in \Cref{fig:20news_semi-synthetic_heatmaps}.
On this semi-synthetic data, NCPD and ONCPD identify topics associated with each subject and accurately indicate the temporal occurrence of each subject, while NMF and LDA learn topics that are prevalent during time slices associated with multiple subjects. 
NCPD and ONCPD learn a single topic for each subject included in the dataset and accurately attribute the highest prevalence to the true underlying topic in each time slice. 
NMF and LDA also learn reasonable topics, including topics corresponding to the longer-lasting ``Space" and ``For Sale" segments.
On this relatively simpler semi-synthetic data, NMF and LDA detect some but not all of the short-lasting topics. 
For example, NMF's learned topic 1 spikes in prevalence during the short-lasting ``Aetheism" and ``Baseball" segments, while LDA accurately detects a short-lasting ``Windows X" related topic.

Both LDA and NMF learn topics that blend multiple document subjects. 
For example, for both NMF and LDA, the most prevalent topic detected during the ``Atheism" time slices is also present during the ``Space" time slices. 
Indeed, we observe that the tensor-based method is able to better detect short-lasting topics and accurately represent them in time.
\begin{figure*}[!t]
    \centering
    \includegraphics[width = \textwidth]{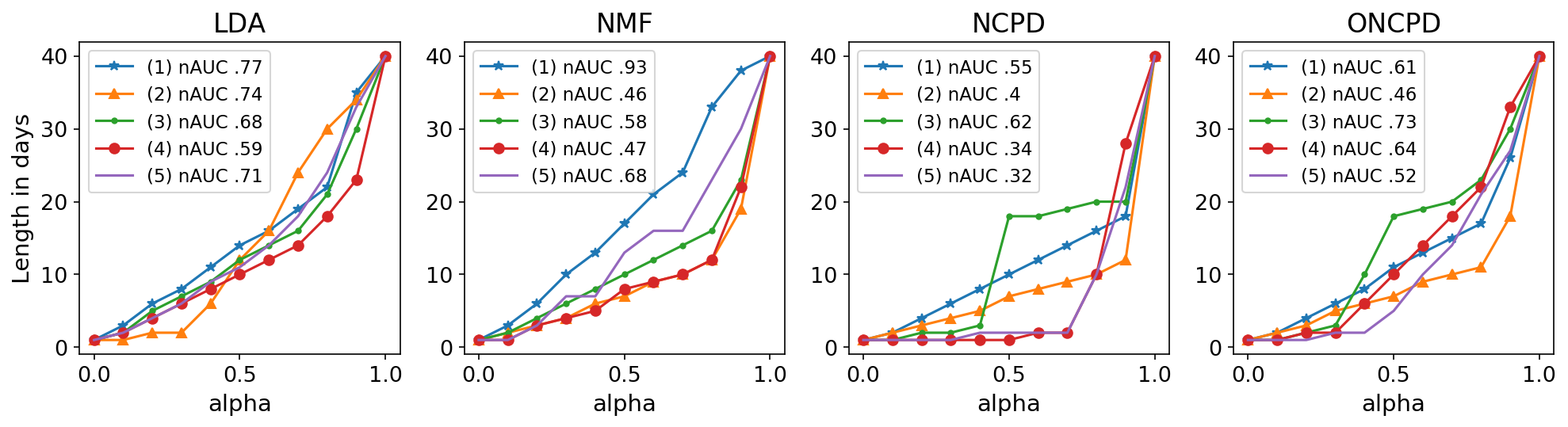}
    \caption{Plot of the $\alpha$-effective lengths of all $5$ topics against $\alpha\in [0,1]$ of the 20news dataset over LDA, NMF, NCPD, and ONCPD methods. The normalized area under the curve (nAUC) is given for each topic in the legend. Smaller nAUC scores indicate shorter-lasting topics, see Section~\ref{sec:topic-lengths}.
    }
    \label{fig:20news_lens}
\end{figure*}

In Figure~\ref{fig:20news_lens}, we track the effective lengths of each of $5$ topics for a range of the values of $\alpha$. The intermediate values of $\alpha \sim 0.5$ can show significant differences in the topic length variability across the methods. We can see that NCPD discovers $2$ topics so that $70\%$ of them appeared within $2$ day period. 
One topic so that its $70\%$ took $8$ days and $2$ more topics that require more than $15$ days for their $70\%$ of the content. In contrast, all the topics discovered by LDA have similar lengths and are generally longer than those discovered by NCPD: for the $70\%$ of the content, all of them require at least a $12$ daytime window. 

With an elbow method, NCPD discovers two short-lasting topics (topics 4 and 5) with the $0.7$-effective length of one day, two topics (topics 2 and 1) of $0.9$-effective lengths of 10 and 18 days, respectively, and one topic (topic 3) of $0.9$-effective length of 20 days that also has $0.4$-effective length of only 2 days (which is, precisely the lengths of these artificially created topics). Choosing $\alpha$ in a shape-agnostic way, with $\alpha\sim 0.5$, we also see that only NCPD method finds 2 short-lasting topics with an effective length of one day and three longer topics with diverse lengths. The legends contain topic numbers referring to the Figure~\ref{fig:20news_semi-synthetic_heatmaps} (for example, topic (3) of the NMF has the top three words ``space, shuttle, nasa"). Additionally, the normalized area under the curve (nAUC) is given for each topic in the legend. It is normalized to be one for a topic uniformly distributed over time. Thus, nAUC shows the persistence of topics by aggregating the $\alpha$-effecting lengths of overall $\alpha$ values in the range from $0$ to $1$. It also shows that LDA tends to find only persistent topics, and NCPD includes more fleeting topics than other methods.

\subsection{News Headlines Dataset Results}
\label{subsec:headlines exp}
\emph{A Million News Headlines} is a dataset containing news headlines published over a period of 17 years sourced from the Australian news source ABC~\cite{DVN/SYBGZL_2018}. The dataset includes noteworthy global events from February 2003 to December 2019 (203 months total) with a focus on Australia.
This dataset combines short-lasting and long-lasting topics, that additionally include one more temporal structure of \emph{periodic} topics (e.g., for seasonal events).
We consider 700 headlines 
randomly selected per month with a total of 142,100 headlines in the entire dataset.
We compute a TFIDF representation for documents, and limit the vocabulary size to 7000, constructing a tensor of shape $(\texttt{Time}\times \texttt{Words}\times \texttt{Docs})=(203\times 7000 \times 700)$.
In these experiments, 20 temporal topics are learned to balance readability and the discovery of relevant events.
\begin{figure*}[!t]
    \centering
    \includegraphics[width = 0.95 \textwidth]{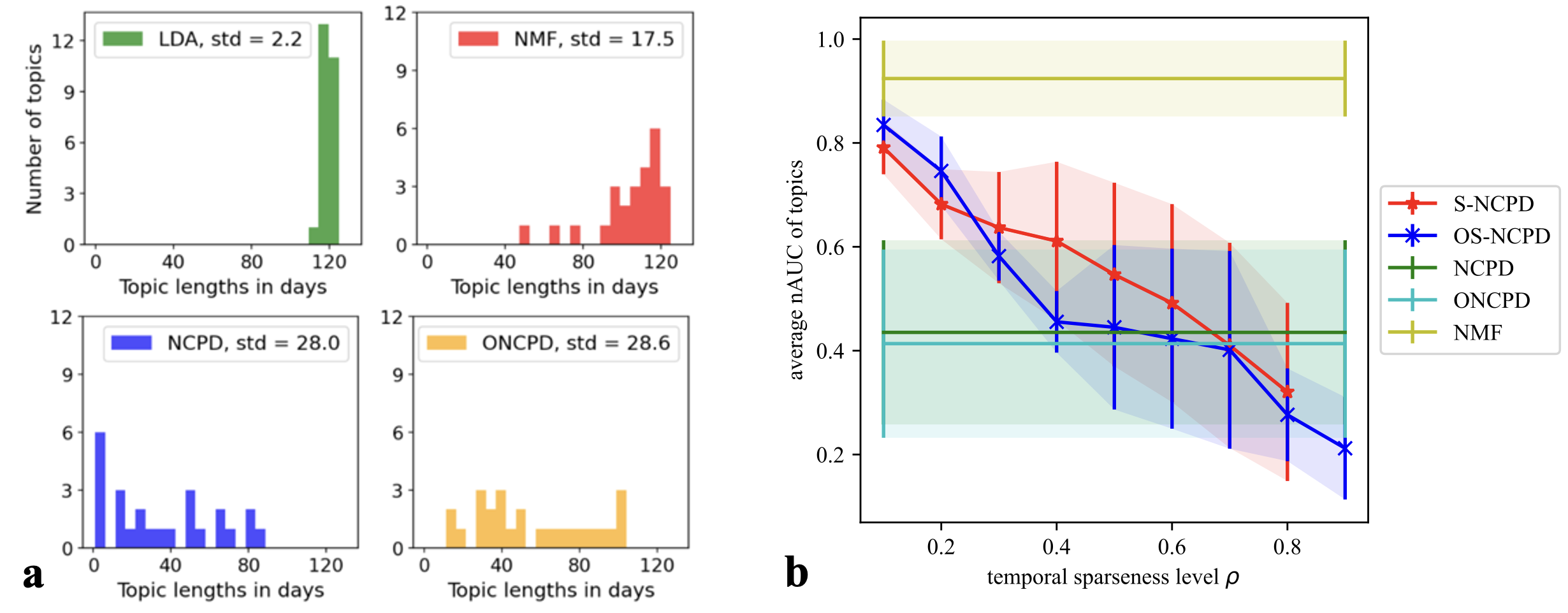}
    \caption{Topic length statistics for Headlines dataset for various methods. 
    \textbf{(a)} Histograms of the $\alpha$-effective lengths with $\alpha=0.6$ of all $25$ topics learned by LDA, NMF, NCPD, and ONCPD. \textbf{(b)} Average nAUC scores (with one standard deviation shown as the shades) of topic lengths vs. temporal sparseness level $\rho$ for S-NCPD, OS-NCPD, NCPD, ONCPD, and NMF. Tensor-based methods are able to learn mixed-length, overall shorter-lasting topics, while the sparseness-constrained methods allow for control of the desired topic length through the sparseness-level-parameter $\rho$. Smaller nAUC scores indicate shorter-lasting topics, see Section~\ref{sec:topic-lengths}.
    }
    \label{fig:nAUC_plot}
\end{figure*}

\begin{figure*}[h]
    \centering
    \includegraphics[width=\textwidth]{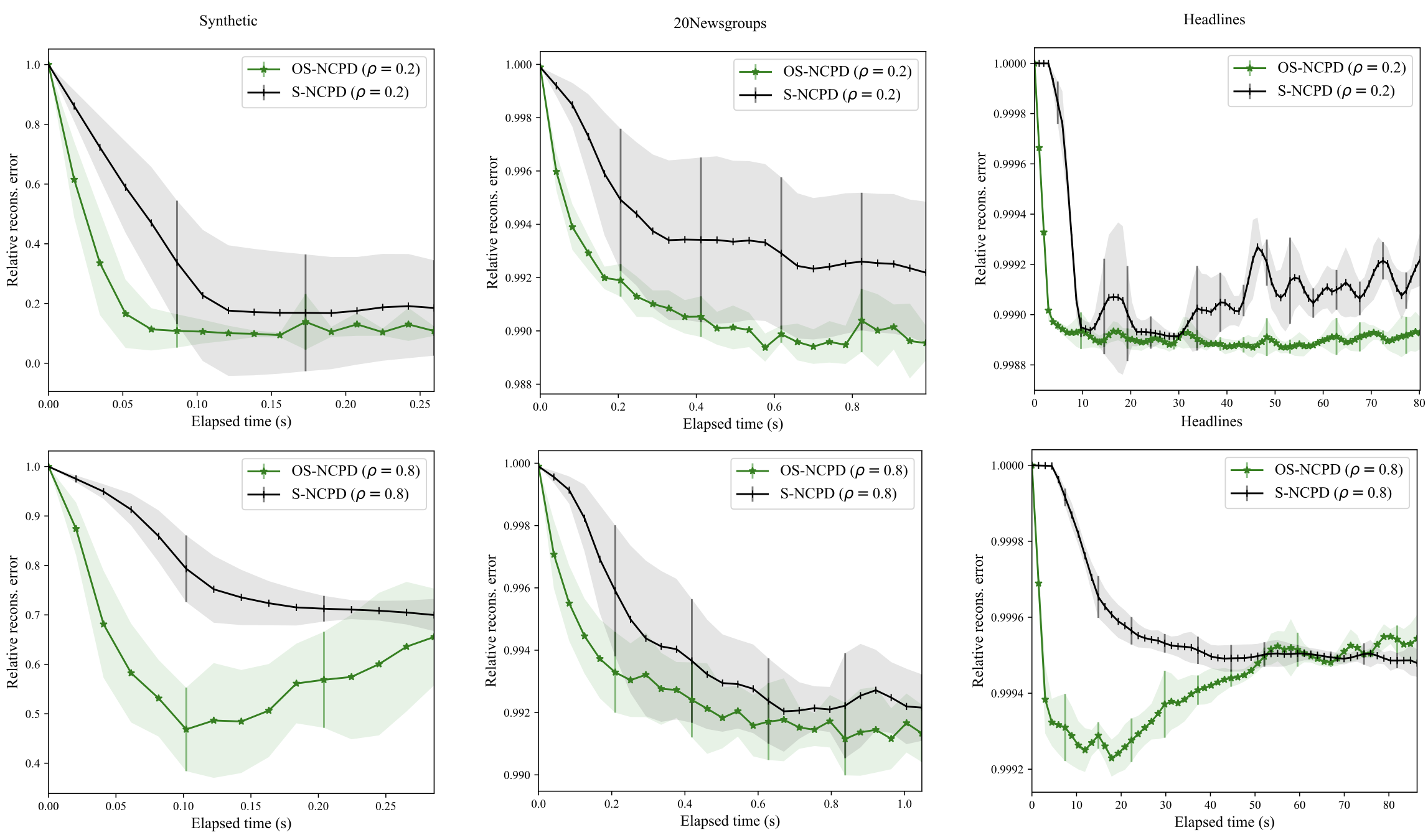}
    \caption{Plots of relative reconstruction error in the time of S-NCPD (black) and OS-NCPD (green) for three datasets: Synthetic tensor (left), 20 Newsgroups (middle), and News Headlines (right). Average and one standard deviation of relative reconstruction errors among ten trials with random initialization are shown.
    }
    \label{fig:benchmark}
\end{figure*}

The upshot of our experiments are summarized in Figures \ref{fig:nAUC_plot} and \ref{fig:benchmark}. We conclude with the following points: 
\begin{enumerate}
    \item (Fig. \ref{fig:nAUC_plot}) LDA and NMF mostly learn long-lasting topics (average nAUC scores $>0.9$) with small variability in topic length (std$<0.15$ nAUC)

    \item (Fig. \ref{fig:nAUC_plot}\textbf{b}) NCPD and ONCPD learn mixed-scale, overall shorter-lasting topics (average nAUC scores $0.4$-$0.42$) with larger variability (std$>0.57$ nAUC) than LDA and NMF. 

    \item (Fig. \ref{fig:nAUC_plot}\textbf{b}) S-NCPD and OS-NCPD learn topics of controlled lengths, where average nAUC scores tend to decay linearly (from 0.8-0.92 to 0.3-0.38) as one increases the sparseness level $\rho$; S-NCPD has larger variability of nAUC scores of the learned topics than OS-NCPD for when targetted to short- or long-lasting topics ($\rho\in (0,0.4)\cup (0.8,1)$); For $\rho\in (0.5,0.7)$, both have large variability (std $\approx 0.4$ nAUC). 

    \item (Fig. \ref{fig:benchmark}) OS-NCPD is significantly more efficient in reducing the reconstruction error than S-NCPD.  
\end{enumerate}

Figure~\ref{fig:nAUC_plot}\textbf{a} demonstrates the histograms of the lengths of all $25$ topics in the Headlines dataset with $\alpha = 0.6$. We can see that LDA produces very similar in length longer topics. Among LDA, NMF, NCPD, and ONCPD, only NCPD is able to pinpoint the six shortest topics with the effective length under $10$ days for $60\%$ of their content (compare with Figure \ref{fig:nmf_ncpd}). Then, ONCPD has the most length variability: sample standard deviations of the lengths of the topics discovered are $2.16$, $17.54$, $28.02$, and $28.6$ respectively for LDA, NMF, NCPD, and ONCPD methods. 

Figure~\ref{fig:nAUC_plot}\textbf{b} plots the `length' of the learned topics measured as the average nAUC scores against the temporal sparseness level $\rho$. It is evident that NMF mostly learns long-lasting topics (average nAUC scores $>0.9$) with small variability in topic length (std$<0.15$ nAUC). On the contrary, NCPD and OCNPD mixed-scale, overall shorter-lasting topics (average nAUC scores $0.4$-$0.42$) with larger variability (std$>0.57$ nAUC) than NMF (see also Fig. \ref{fig:nmf_ncpd} and \ref{fig:oncpd}). While one cannot control the temporal structure of topics to be learned via these methods, we see that the average topic lengths for S-NCPD and OS-NCPD decay linearly in the temporal sparseness level $\rho$. There, the average nAUC scores tend to decay linearly (from 0.8 to 0.2) as one increases the sparseness level $\rho$. As for the variability of nAUC scores (i.e., the range of temporal scales of the learned topics), S-NCPD has larger variability of nAUC scores of the learned topics than OS-NCPD for when targetted to short- or long-lasting topics ($\rho\in (0,0.4)\cup (0.8,1)$); For $\rho\in (0.5,0.7)$, both have large variability (std $\approx 0.4$ nAUC), resembling NCPD and ONPCD (see Fig. \ref{fig:oncpd} left). 

Figure \ref{fig:benchmark} shows relative reconstruction error in time of S-NCPD and OS-NCPD with $\rho\in \{0.2, 0.8\}$ for synthetic tensor of shape $(100\times 200 \times 300)$, the semi-synthetic 20 Newsgroups tensor (Section \ref{subsec:semi synth}), and the News Headlines tensor. The average relative reconstruction errors are shown with one standard deviation in shades. In all experiments, we see that OS-NCPD decreases the objective value much faster than S-NCPD. Since we use a heuristic solver (Alg. \ref{algorithm:SNLS}) for solving the sparsity-constrained nonnegative least squares, the objective value can fluctuate as the algorithms proceed. This is in contrast to the monotone decrease in the objective value for NCPD and ONCPD observed in \cite[Fig. 2]{lyu2022online}.

We give a more detailed discussion through Figures \ref{fig:LDA_headlines}, \ref{fig:nmf_ncpd}, \ref{fig:sncpd}, \ref{fig:osncpd}, and \ref{fig:oncpd}.

\subsubsection{NMF and LDA: Learn mostly long-lasting topics}
In order to use NMF to detect topics and their time evolution, we may preprocess the 3D tensor into a $\texttt{Time}\times \texttt{Words}$ tensor in the following two ways: (1) unfold the 3D tensor so that the resulting 2D tensor is a concatenation of the word frequency vectors of individual documents (total of 700*203); (2) average the word frequency vectors for all 700 documents within each month into a single word frequency vector.  Applying NMF on (1) does not seem to detect topics of clear temporal structure, as shown in Figure \ref{fig:nmf_ncpd}. The prevalence of the topics (measured by nAUC scores) shown in Figure \ref{fig:nmf_ncpd} indicates that NMF can only learn long-lasting topics (of nAUC scores close to one). 

\begin{figure*}[h]
 \includegraphics[width=0.99\columnwidth]{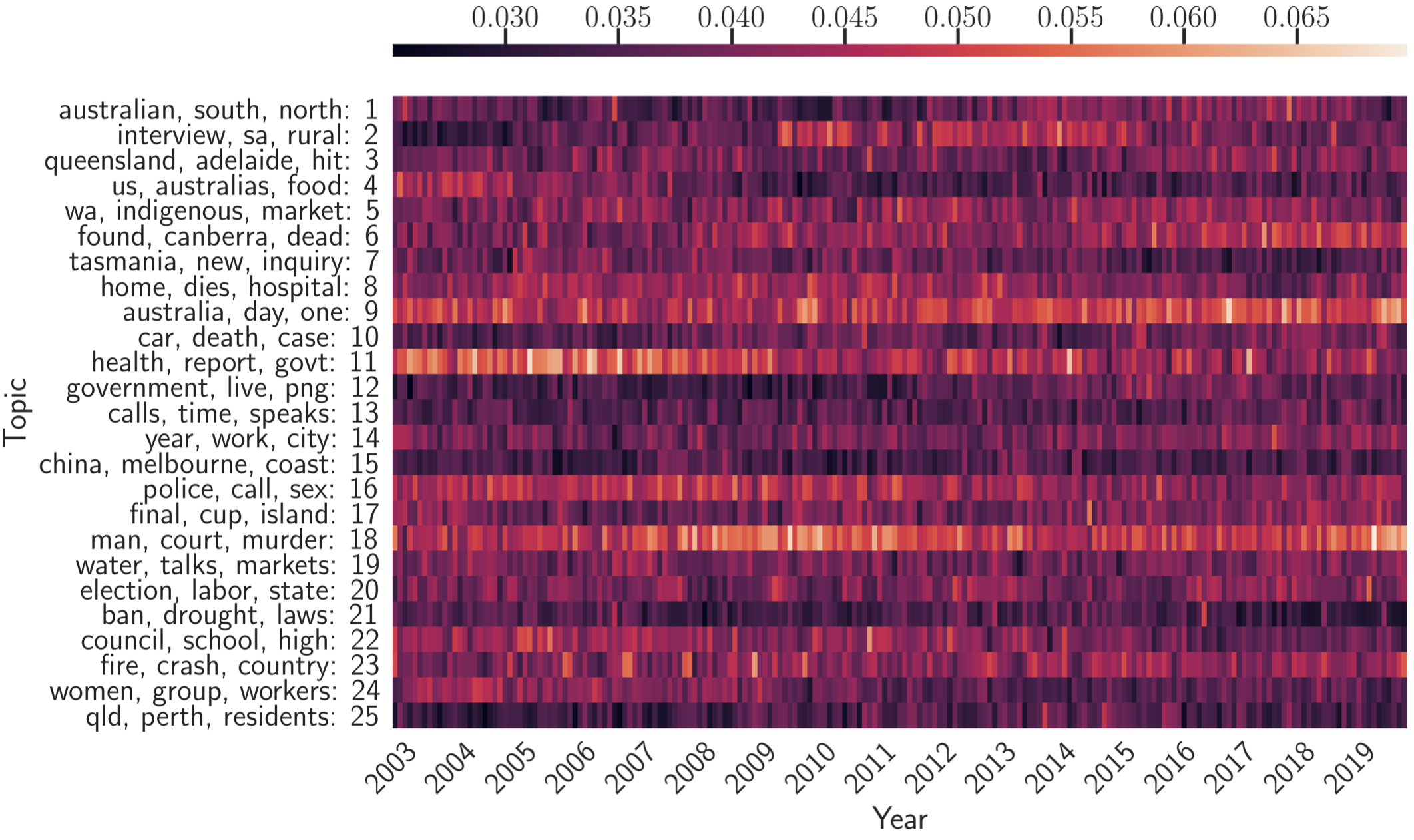}
    \caption{25 temporal topics are learned from ABC News Headlines dataset 
    (a tensor of shape $(\texttt{Time}\times \texttt{Words}\times \texttt{Docs})=(203\times 7000 \times 700)$) via LDA We show three top keywords for each topic and its time evolution as a heatmap, indexed by years. The heapmaps are normalized such that the sum of the weights over the whole topic at each time period equals one.}
    \label{fig:LDA_headlines}
\end{figure*}

\begin{figure*}[h]
 \includegraphics[width=0.99\columnwidth]{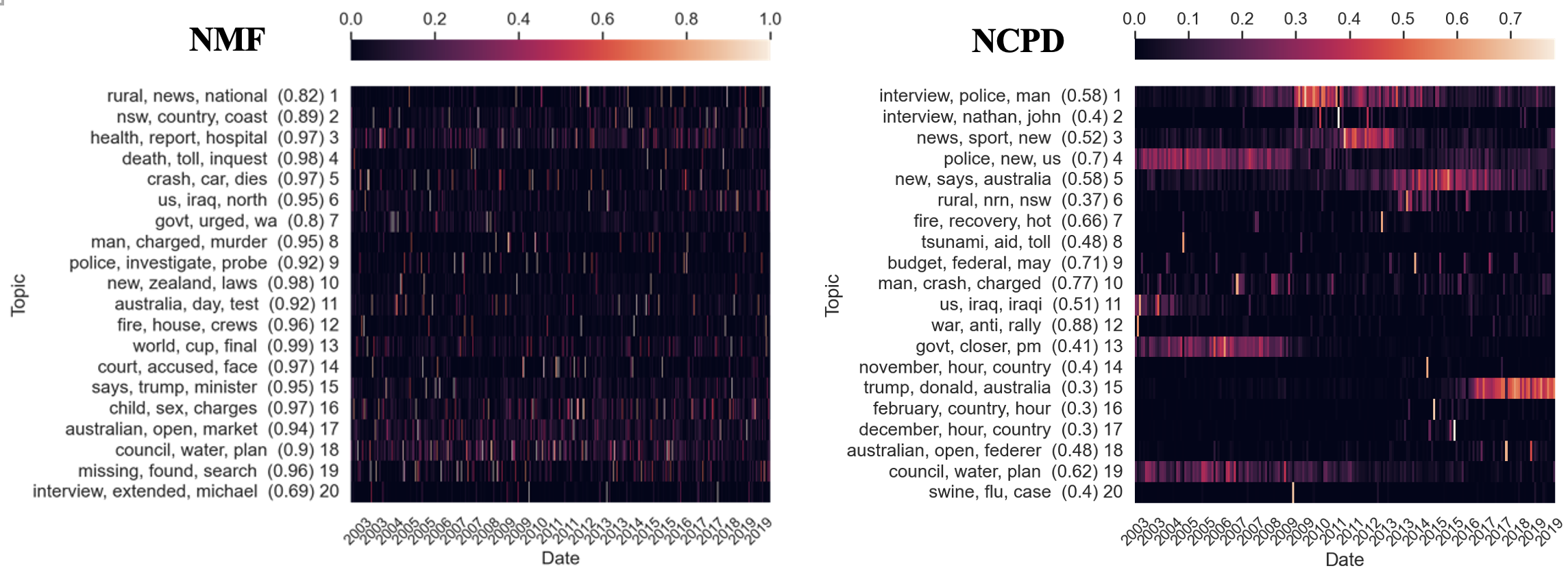}
    \caption{20 temporal topics are learned from ABC News Headlines dataset 
    (a tensor of shape $(\texttt{Time}\times \texttt{Words}\times \texttt{Docs})=(203\times 7000 \times 700)$) via NMF and NCPD. The prevalence (measured by nAUC scores) of each topic is shown in parentheses. We show three top keywords for each topic and its time evolution as a heatmap, indexed by years. The heapmaps are normalized so that the rows sum to one}
    \label{fig:nmf_ncpd}
\end{figure*}

Preprocessing (2) suffers from merging many documents of potentially distinct topics into one, so one can expect the topics detected by NMF would mix keywords from different topics. We omitted a similar plot for this experiment. Also, LDA was only able to detect topics whose prevalence spans the entire temporal horizon (see Fig. \ref{fig:LDA_headlines}). In comparison to the semi-synthetic data in Figure \ref{fig:20news_semi-synthetic_heatmaps}, we find that LDA is not effective in detecting short-lasting and periodic topics from real data. 

\subsubsection{NCPD and ONCPD: Learn mixed-scale, overall shorter-lasting topics}
\begin{figure*}[h]
 \includegraphics[width=0.99\columnwidth]{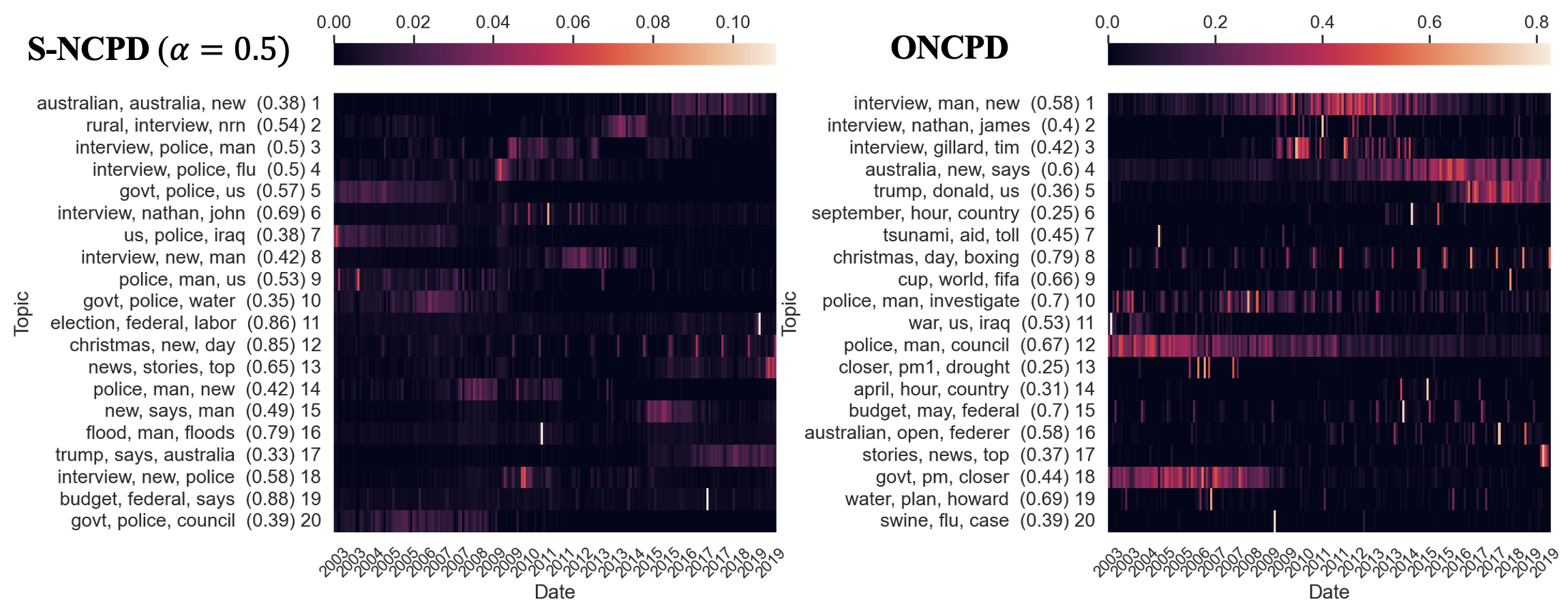}
    \caption{20 temporal topics are learned from ABC News Headlines dataset 
    (a tensor of shape $(\texttt{Time}\times \texttt{Words}\times \texttt{Docs})=(203\times 7000 \times 700)$) via S-NCPD (with $\rho=0.5$) and ONCPD. We show three top keywords for each topic and its time evolution as a heatmap, indexed by years. The prevalence (measured by nAUC scores) of each topic is shown in parentheses. The heapmaps are normalized so that the rows sum to one}
    \label{fig:oncpd}
\end{figure*}

We observe in Figure \ref{fig:nmf_ncpd} that (standard) NCPD automatically detects short-lasting, periodic (e.g., topic 20 on "swine", "flu", and "case"), and long-lasting topics (e.g., topic 4 on "police", "news", and "us"). In particular, as seen in Figure \ref{fig:nmf_ncpd}, NCPD is able to learn topics with small nAUC scores (e.g., nAUC$=0.4$ for topic 20) as well as large nAUC scores (e.g., nAUC$=0.88$ for topic 12). From the keywords of these topics, we observe relatively more cohesive topics that align with real-world events. E.g., topic 18 (``Australian" ``open", ``Federer"), topic 9 ("budget", "federal", "May"). The topics learned by ONCPD share very similar characteristics to the ones learned by NCPD (see Fig. \ref{fig:oncpd}). 

Compared to the NMF experiment in Figure \ref{fig:nmf_ncpd}, NCPD can detect meaningful topics with a clear temporal structure. The key difference is that NCPD processes the thrid-order tensor data at once, where multiple documents within the same temporal documents (specifically, 708 documents in our Headlines dataset) are considered to be simultaneous while keeping different documents separate so that no two documents of distinct topics are merged in the pre-processing stage (as in NMF pre-processing scheme (2)). We mention that while it is possible to use the final reconstruction error of NCPD to assess the goodness of the overall factorization, computing the reconstruction error in this case is prohibitively expensive as it involves processing 20 tensors (one for each topic) of shape $(203\times 7000\times 700)$. 

\subsubsection{S-NCPD and OS-NCPD: Controlled temporal structure}

 \begin{figure*}[h]
 \includegraphics[width=0.99\columnwidth]{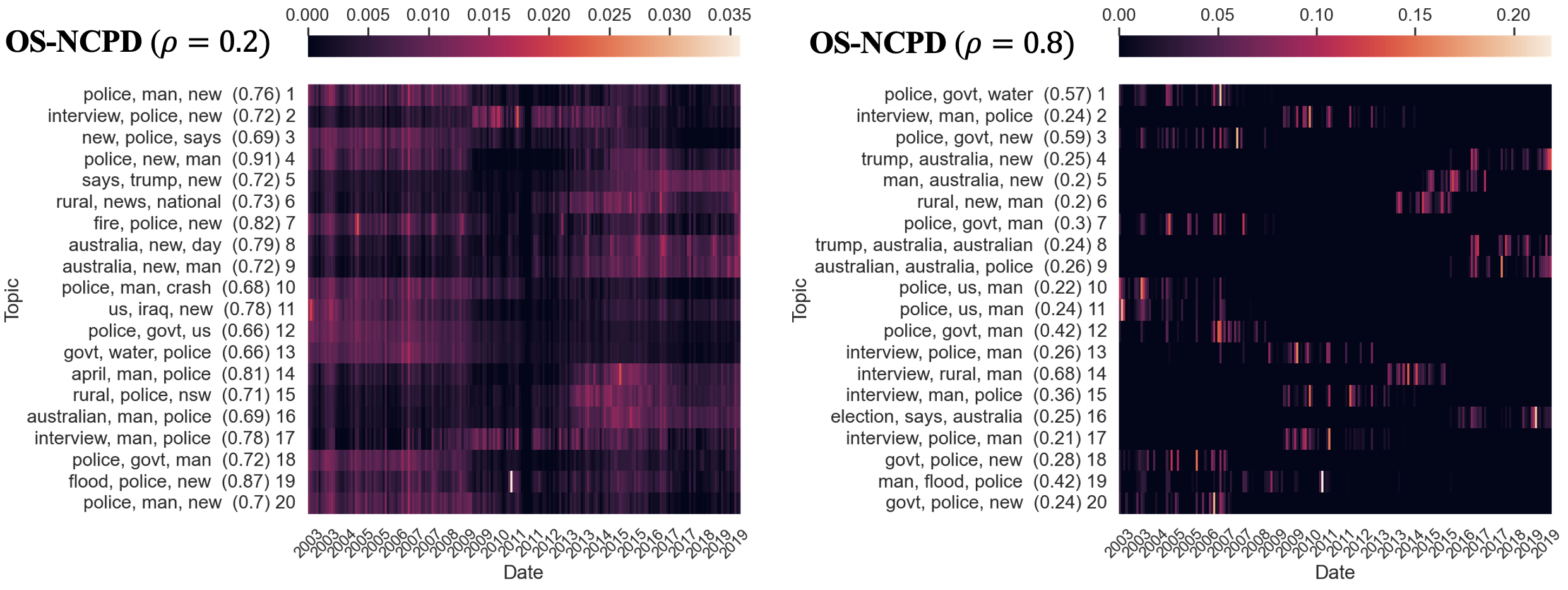}
    \caption{20 temporal topics are learned from ABC News Headlines dataset 
    (a tensor of shape $(\texttt{Time}\times \texttt{Words}\times \texttt{Docs})=(203\times 7000 \times 700)$) via OS-NCPD with various temporal sparseness levels $\alpha$. We show three top keywords for each topic and its time evolution as a heatmap, indexed by years. The prevalence (measured by nAUC scores) of each topic is shown in parentheses. The heapmaps are normalized so that the rows sum to one}
    \label{fig:osncpd}
\end{figure*}

 \begin{figure*}[h]
 \includegraphics[width=0.99\columnwidth]{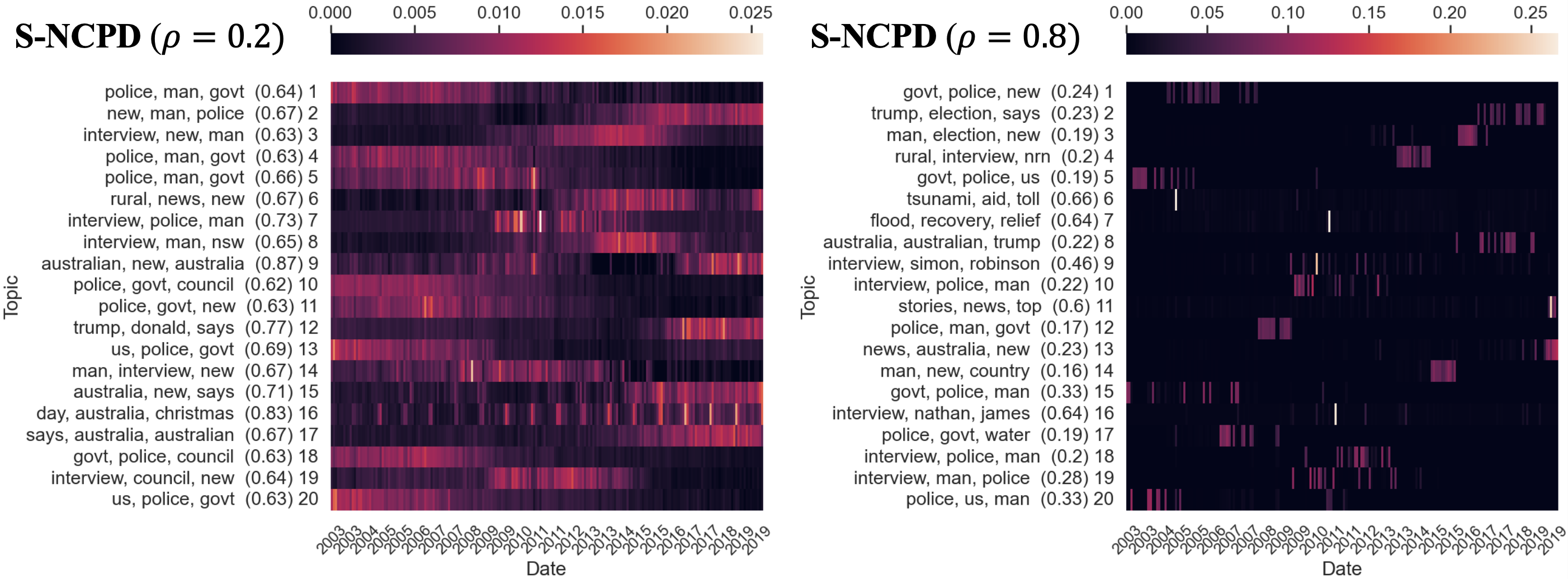}
    \caption{20 temporal topics are learned from ABC News Headlines dataset 
    (a tensor of shape $(\texttt{Time}\times \texttt{Words}\times \texttt{Docs})=(203\times 7000 \times 700)$) via S-NCPD with various temporal sparseness levels $\rho=0.2$ and $0.8$. We show three top keywords for each topic and its time evolution as a heatmap, indexed by years. The prevalence (measured by nAUC scores) of each topic is shown in parentheses. The heapmaps are normalized so that the rows sum to one}
    \label{fig:sncpd}
\end{figure*}

While we see that NCPD can detect topics of various temporal characteristics, it would be beneficial to have methods for actively controlling the desired length of topics. We proposed S-NCPD and OS-NCPD for this purpose. If we use sparseness level $\alpha=0.2$ for S-NCPD as in Figure \ref{fig:sncpd}, it would restrict NCPD to learn topics whose time evolution (i.e., the corresponding columns in the $\texttt{time}\times \texttt{topic}$ factor $\mC$ matrix) has sparseness level $0.2$, so it is rather evenly distributed over the entire time horizon. On the other hand, using $\alpha=0.8$ as in Figure \ref{fig:sncpd} now promotes learning only topics with much shorter prevalence. This additional temporal sparseness restriction in general results in fewer distinct topics compared to vanilla NCPD but could uncover new topics that were not detected by vanilla NCPD. For instance, with
sparseness level 0.8 (Figure \ref{fig:sncpd}), we uncover a topic (topic 7: ``flood", ``recovery", ``relief") not readily discovered by vanilla NCPD with rank 20 by the top keywords. A similar discussion as above also applies to OS-NCPD (see Fig. \ref{fig:osncpd}). However, there are notable differences in the standard deviation of the nAUC scores of the topics learned by S-NCPD and OS-NCPD. When $\rho$ is tuned so that either short-lasting or long-lasting topics are targeted, OS-NCPD tends to result in a smaller variation of the nAUC scores than S-NCPD (see Fig. \ref{fig:nAUC_plot}\textbf{b}). 

\subsubsection{Computational efficiency of OS-NCPD over S-NCPD}\label{online_section}

An obvious disadvantage of S-NCPD is the computational cost of finding the sparsity-constrained nonnegative CP decomposition and the memory required to store the whole tensor. We show that OS-NCPD provides a viable alternative to the S-NCPD method for the limited computational resources.

We compare the performance of S-NCPD \eqref{alg:SNCPD} and OS-NCPD \eqref{alg:OS_NCPD} on three datasets (synthetic tensor, semi-synthetic 20 Newsgroup, and News Headlines) in terms of the relative reconstruction error at various temporal sparseness levels. For each dataset, we run each of the algorithms with rank $5$ ten times with randomly initialized factor matrices with independent entries sampled uniformly from the interval $[0,1]$. In \Cref{fig:benchmark}, the average of reconstruction errors (computed by \eqref{eq:NCPD_reconstruction})
with 1 standard deviation are shown by the solid lines and shaded regions of respective colors. 

 OS-NCPD works with smaller data tensors of size $(\mathtt{words}, \mathtt{batch}', \mathtt{time})$ (see also the discussion in Subsection \ref{sec:ONCPD}), where we may take $\mathtt{batch}'$ arbitrarily smaller than the actual number of documents $\mathtt{batch}$ in the original data tensor. From this, one can expect that the OS-NCPD is more computationally efficient than the OS-NCPD algorithm.
Indeed, in Figure \ref{fig:benchmark}, we see that OS-NCPD is able to decrease the reconstruction error much more rapidly than the standard S-NCPD, although given enough time and computational budget, OS-NCPD may eventually end up with a smaller reconstruction error than OS-NCPD as in the 20 Newsgroups data in Figure \ref{fig:benchmark}.

Also, it is important to reiterate that such a computational gain in using OS-NCPD in dynamic topic modeling does not necessarily entail a compromise in the ability of NCPD to learn a variety of short-term and long-term topics (e.g., in the News Headlines). 

\section{Conclusion and Future Work}\label{sec:conclusion}

We demonstrate nonnegative CANDECOMP/PARAFAC decomposition (NCPD) as a powerful dynamic topic modeling technique capable of detecting short-lasting and periodic topics along with long-lasting topics in dynamic text datasets. In order to overcome the lack of controllability of topic lengths in NCPD, we proposed two new methods that can actively control the lengths of topics through an additional sparseness constraint. We propose both the offline (S-NCPD) and online (OS-NCPD) versions of such methods. We discuss and compare the temporal topic patterns learned through each of these methods. We propose different ways to measure the lengths of the discovered topics and validate the ability of tensor methods to discover short-term topics quantitatively. We observe that both S-NCPD and OS-NCPD extract fewer distinct, but potentially new topics depending on the temporal sparseness parameter, where the average topic lengths decrease linearly as we increase that parameter. For large datasets, OS-NCPD serves as a viable alternative for learning topics and their temporal patterns, retaining the ability to detect controlled short-lasting topics. 

Among the natural future directions of the current work, is improving the efficiency of nonnegative tensor decompositions, e.g., by employing geometry-preserving tensor dimension reduction techniques (such as, \cite{iwen2021lower}), running NCPD fitting algorithms on a compressed tensor, and subsequent recovery of the topics from their compressed representation.
Additionally, it is interesting to study the prominence evolution of a particular topic with respect to the others via tensor extensions of the recently proposed GuidedNMF algorithm \cite{vendrow2021guided}.
We also aim to study the relation between the sparseness level in the temporal component of the tensor and the rank of the decomposition.

Finally, the proposed methods S-NCPD and OS-NCPD are not specific for a particular type of data. Finding the topics, or clusters of data, with controlled localization properties would be important for various applications (not considered in this paper) where non-negative low-rank matrix and tensor methods are extensively employed, including the text data coming from multiple sources \cite{vendrow2022generalized}, image analysis \cite{lee1999learning, kumar2018hyperspectral, he2014semi}, or computational biology \cite{brunet2004metagenes, kim2007sparse, alexandrov2013deciphering}. 

\section*{Funding}
 DM, DN, and ER were partially funded by NSF DMS 2011140. ER was also partially funded by NSF DMS 2309685 and NSF DMS 2108479. HL was partially funded by NSF DMS 2206296 and DMS 2010035.

\section*{Acknowledgments}
We thank Jacob Moorman for his contributions to the code used. ER is also thankful to Maria Avdeeva for very useful discussions.

\bibliography{bib}   

\end{document}